\documentclass[aps,nofootinbib,showpacs,twocolumn,floatfix,superscriptaddress]{revtex4}
\usepackage{graphicx}
\usepackage{epsf,amsfonts,amssymb,amsbsy}
\usepackage[mathscr]{eucal}
\begin{document}
\title{Properties of potential modelling three benchmarks:\\ the cosmological constant, inflation
and three generations}
\author{V.V.Kiselev}
 \affiliation{Russian State Research
Center ``Institute for High
Energy Physics'', 
Pobeda 1, Protvino, Moscow Region, 142281, Russia\\ Fax:
+7-4967-742824}
 \affiliation{Moscow Institute of Physics and
Technology, Institutskii per. 9, Dolgoprudnyi, Moscow Region,
141700, Russia}
\author{S.A.Timofeev}
\affiliation{Moscow Institute of Physics and Technology,
Institutskii per. 9, Dolgoprudnyi, Moscow Region, 141700, Russia}
\begin{abstract}
We argue for a model of low-energy correction to the inflationary
potential as caused by the gauge-mediated breaking down the
supersymmetry at the scale of
$\mu_{\mbox{\footnotesize\textsc{x}}}\sim 10^4$ GeV, that provides
us with the seesaw mechanism of thin domain wall fluctuations in
the flat vacuum. The fluctuations are responsible for the vacuum
with the cosmological constant at the scale of $\mu_\Lambda\sim
10^{-2}$ eV suppressed by the Planckian mass $m_\mathtt{Pl}$ via
$\mu_\Lambda\sim\mu_{\mbox{\footnotesize\textsc{x}}}^2/m_\mathtt{Pl}$.
The appropriate vacuum state is occupied after the inflation with
quartic coupling constant
$\lambda\sim\mu_{\mbox{\footnotesize\textsc{x}}}/m_\mathtt{Pl}\sim
10^{-14}$ inherently related with the bare mass scale of
\mbox{$\widetilde
m\sim\sqrt{\mu_{\mbox{\footnotesize\textsc{x}}}m_\mathtt{Pl}}\sim
10^{12}$ GeV} determining the thickness of domain walls $\delta
r\sim1/\widetilde m$. Such the parameters of potential are
still marginally consistent with the observed inhomogeneity of matter density in
the Universe. The inflationary evolution suggests the vacuum
structure compatible with three fermionic generations of matter as
well as with observed hierarchies of masses and mixing in the
Standard Model.
\end{abstract}
\maketitle

\section{Introduction}

Recently the seesaw mechanism for the mixing of two virtual
vacuum-levels due to fluctuations described by thin domain walls,
has been explored in order to derive the natural scale of
cosmological constant \cite{Weinberg-RMP} in terms of supersymmetry
(SUSY) breaking scale $\mu_{\mbox{\footnotesize\textsc{x}}}$ and
Planck mass defined by the Newton constant $G$ in gravitation,
$m_\mathtt{Pl}=1/\sqrt{G}$, so that the vacuum state of de Sitter
spacetime (dS) constitutes the stationary level composed by the
superposition of flat and Anti-de Sitter (AdS) vacua, and it has got
the constant energy density $\rho_\Lambda= \mu_\Lambda^4$ at
$\mu_\Lambda\sim\mu_{\mbox{\footnotesize\textsc{x}}}^2/m_\mathtt{Pl}$
\cite{KT1}. Then, the modern value of cosmological constant
\cite{Weinberg-RMP,WMAP5-1,WMAP5-2,WMAP7,BAO,Riess:2004nr,Riess:2006fw,
Astier:2005qq,WoodVasey:2007jb}
gives $\mu_\Lambda\approx 0.25\cdot 10^{-2}$ eV, hence,
$\mu_{\mbox{\footnotesize\textsc{x}}}\sim10^4$ GeV, i.e. the low
scale of SUSY breaking down. The virtual flat vacuum corresponds to
a scalar field positioned in the local minimum of its potential with
zero energy provided by the exact SUSY. The virtual AdS vacuum-state
is described by the field positioned at the local minimum of primary
potential with positive energy, which breaks down SUSY, while the
supergravity contribution linear in $G$ \cite{Weinberg-VIII} gives
the term providing the negative overall sign of cosmological
constant due to the dominant energy density of zero-point
quantum-field modes with masses
$m\sim\mu_{\mbox{\footnotesize\textsc{x}}}$. The virtual flat vacuum
does not decay to the AdS one\footnote{If the gravity is switched
off, the false vacuum decay, see \cite{false-decay}.} due to
stabilization effect of gravity \cite{CdL,SW}, but it suffers from
fluctuations in the form of spherically symmetric AdS-bubbles
surrounded by domain walls. Then, due to such the specific seesaw
mechanism,  one of two true
stationary states is the dS-vacuum. Note, that the low energy
contributions due to such phenomena like the electroweak symmetry
breaking down or condensates in QCD can modify the vacuum energy of
initial AdS-state with broken SUSY, only, while zero energy of flat
state is preserved by exact SUSY. However, the numerical value of
low energy condensates is negligible in comparison with the dominant
term coming from the SUSY breaking itself due to the appropriate
hierarchy of relevant scales. Similarly, any dynamics at energies
higher than the scale of SUSY breaking down cannot disturb the
vacuum energy, since such the dynamics is supersymmetric.

The quite general idea of incorporating the seesaw mechanism for the
derivation of naturally small cosmological constant from the Planck
mass and SUSY breaking scale is not originally new itself. Such the
point of view was presented in scientific e-folklore as discussion
in blogs\footnote{See the following blog sites
http://cosmicvariance.com/2005/12/\\
05/duff-on-susskind/\#comment-8629 (on Dec 6, 2005) and
http://motls.blogspot.com/2005/12/cosmological-constant-seesaw.html
(on Dec 19, 2005).}, for instance. In addition, in the framework of
supergravity G.~Chalmers argued for the relevant suppression of
cosmological constant in \cite{Chalmers}. However, the idea becomes
more actual, when it is realized in terms of reasonable model.
M.~McGuigan has modified the Wheeler-DeWitt equation in order to
switch on a coupling between two sectors characterized by the Planck
scale and SUSY, correspondingly. So, the seesaw mechanism has been
involved into the gravity, and the cosmological constant of natural
scale has been generated \cite{Grav-seesaw}. The other way has been
formulated in our approach invented in \cite{KT1}.

In \cite{KT2} we have constructed a model of potential, which has
allowed us to investigate the scale parameters in the problem within
the suggested  approach. So, we have established the
following general features:
\begin{itemize}
    \item thin domain walls correspond to the low-scale SUSY
    breaking down due to the gauge mediation, when the distance between the
    extremal positions of scalar field takes sub-Planckian values, while
    \item thick domain walls are related with the gravity-mediated
    SUSY breaking down at high energies
    $\mu_{\mbox{\footnotesize\textsc{x}}}\sim 10^{12-13}$ GeV and
    super-Planckian values of field increment between the extremals.
\end{itemize}
This potential is suitable also for demonstrating the origin of
fermion generations observed in the Standard model. Indeed, the
non-trivial vacuum structure is described by a superposition of
initial states with definite masses of fermions. The superposition
can be represented by two-dimensional (2D) column, say. Therefore,
it is natural that a $2\raisebox{1pt}{$\scriptstyle\times$} 2$
mass-matrix is involved for the fermion states, i.e. two
generations appear. One can easily introduce three generations by
considering the superposition of three initial vacuum-states: two
flat levels and single AdS-vacuum in the model discussed. However,
such the vacuum structure does not answer the question, why we are
living in the vacuum we have got, since all of three stationary
superpositions can be occupied, but one of them, at least, is the
AdS-state with a huge negative energy density irrelevant to the
astronomical observations. Moreover, it is the flat vacuum
suffering from the domain-wall fluctuations. Therefore, evolving
the scalar field to the position of flat vacuum will incorporate
the mixing with the AdS-vacuum only, and the evolution will not
see the mixing with another flat vacuum even through the
AdS-states, since beyond the domain wall the boundary condition at
spatial infinity remains unchanged, i.e. ascribed to the first
flat vacuum. This fact does not influence the analysis as concerns
for scaling properties of potential, of course. However, the
potential needs a modification compatible with the scalar-field
evolution during the Universe expansion in order to get the
natural reason for the living in the dS-vacuum. In addition, the
potential of \cite{KT2} suggests a kind of fine tuning, because of
its two flat vacua with the coinciding zero vacuum-energy at
different values of scalar field.

The direction of modification is clear: one should get a potential
with a single flat vacuum and a couple of AdS-vacua, which energy
densities $\rho=-\rho_{\mbox{\footnotesize\textsc{x}}}^{\pm}$ do
not coincide, in general, but
$\rho_{\mbox{\footnotesize\textsc{x}}}^{\pm}$ take values of the
same order of magnitude. Then, the flat vacuum will fluctuate due
to bubbles of both AdS-vacua, and the Universe will get the
observed cosmological constant in the dS-state, if the evolution
will drive it to the field-position in the flat vacuum.

In Section \ref{S-II} we present the model of potential satisfying
all requirements listed above. The potential is composed of
several contributions. The first term is the bare quadratic
potential, which generates the second contribution being the
quartic term due to the supergravity correction linear in Newton
constant $G$. The third term modifies the potential at low
energies due to the modelled contribution by zero-point modes, so
that it has got the form with the flat local minimum and two AdS
minima. The barrier between the flat and AdS minima is tuned in
order to produce the thin domain wall with the thickness given by
the inverse bare mass in the first term mentioned above. Moreover,
the behavior of potential at $\phi\to 0$ is well approximated by
the quartic term with the coupling constant of the same order of
magnitude at large fields. Then, we can evaluate the bare mass
$m_\mathrm{bare}\sim 10^{12}$ GeV and quartic coupling
$\lambda\sim 10^{-14}$.

The vacuum structure formed due to the mixing between the initial
flat and AdS vacua because of domain wall fluctuations, is
described in Section \ref{S-III}. Then, the stationary dS-vacuum
state appears due to the fluctuations in vicinity of flat vacuum.

In Section \ref{S-IV} we study the inflation\footnote{See modern
review in \cite{inflation}.}
\cite{i-Guth,i-Linde,i-Albrecht+Steinhardt,i-Linde2} governed by
the scalar field with the quadric and quartic self-couplings. The
field evolution corresponds to the dynamical system possessing the
properties of parametric attractor: the kinetic and potential terms
rapidly reach stable critical points, which have got a slow
driftage with the growth of e-folding in the scale factor of
accelerated expansion. Then, the quasi-attractor provides us with
the tool to quantify the inhomogeneity generated by the quantum
fluctuations of scalar field during the inflation. The data
signalize for the dominance of quadratic term in the potential. We
give a numerical constraint for the dominance and discuss the
conditions of its realization in consistency with the
observations. After the inflation, the scalar field enters the
stage of preheating due to the tachyonic mechanism: the potential
barrier  
generates a negative square of effective mass, that results in the
scalar field decay to massless quanta in vicinity of flat vacuum.
The preheating should take place at the temperature
$T_\mathrm{preh.}\sim 10^9$ GeV determined by the barrier height.
Thus, the universe evolution drives to the dS-vacuum state.

Note that the review on the relation of inflation to the particle
physics and on the mechanism of preheating and thermalization of Universe after
the inflation\footnote{Realistic models of low-energy inflation taking into
account of constraints following from the primordial nucleosynthesis (Big Bang
nucleosynthesis, anisotrophy of cosmic background radiation and inhomogeinity
of matter density in the large scale structure of Universe, are presented in 
\cite{Allahverdi:2006iq,Allahverdi:2006cx,Allahverdi:2006we}, wherein a
supersymmetric version of Standard Model for the particle interactions is
studied with the use of flat directions in superpotentials.} 
can be found in \cite{Mazumdar:2010sa}.

Section \ref{S-V} is devoted to the analysis of textures in the
mass matrices for three generations of charged fermions as 
caused by the vacuum structure. 
We show that the hierarchies in masses and mixings of
charged weak currents as well as a fine violation of combined
inversion of charge and space can be natural for the fermions of
observable sector, while the hidden sector responsible for the
SUSY breaking down can remain heavy. The picture similar to the
hidden sector could take place for the sfermion partners of
observable fermions.

In Section \ref{S-VI} we summarize our results and discuss
problems of the model offered.

\section{The potential model\label{S-II}}
For the sake of simplicity, we ascribe the energy density of
initial vacuum-state $\rho=-\rho_{\mbox{\footnotesize\textsc{x}}}$
to the effective contribution of single fermionic zero-point
quantum-field mode
\begin{equation}\label{z7}
\begin{array}{rl}
    \hat \rho(\mathscr{M}) =&\displaystyle
    \frac{1}{2}\int\limits_{\mathscr{M}}^{\mu_{\mbox{\footnotesize\textsc{x}}}}
    \frac{k^2{\rm d}k}{(2\pi)^3}\;\sqrt{
    k^2+m^2}\int{\rm d}\Omega\\[6mm]
    =& \displaystyle\frac{2}{(16\pi)^2}\;m^4\,(\sinh
    4y-4y)\Big|_{y(\mathscr{M})}^{y(\mu_{\mbox{\footnotesize\textsc{x}}})}
\end{array}
\end{equation}
with
$$
    y(\mu)=\mbox{arcsinh}\frac{\mu}{m}=\ln\left(
    \frac{\mu}{m}+\sqrt{\frac{\mu^2}{m^2}+1}\;
    \right),
$$
and $\mathscr{M}\in[0,\mu_{\mbox{\footnotesize\textsc{x}}}]$. At
$\mathscr{M}=\mu_{\mbox{\footnotesize\textsc{x}}}$, SUSY is exact,
while at $\mathscr{M}=0$ we get
$\rho_{\mbox{\footnotesize\textsc{x}}}=\hat\rho(0)$ and SUSY is
broken down. Here, $\mathscr{M}=\mathscr{M}(\phi)$ is actually
expressed in terms of canonic scalar field $\phi$ representing the
component of chiral superfield. Below we put $\phi$ to be real. 
This constraint is introduced by two following  requirements:
\begin{description}
 \item[i)] Phenomenologically, the inflation is well described by a single real
scalar field called ``the inflaton'', which is ascribed to the real component
of scalar field in the chiral superfield in the framework of our model.
\item[ii)] During the inflation, the supersymmtery is suggested broken
down, so that the imaginary part of scalar
component of chiral superfield acquires a mass of Planckian scale, that makes
its dynamics irrelevant (or simply frozen) at the inflationary stage under
study. In that case, the shift symmetry of superpotential becomes invalid, and
it remains beyond the scope of problem in touch.
 \end{description}

The complete actual potential energy including linear corrections in $G$
from supergravity, has the form\footnote{See, for instance,
\cite{Weinberg-VIII}.}
\begin{equation}\label{mg2}
    \begin{array}{cl}
      U(\phi)= &\displaystyle
      V(\phi)-24\pi\,G\left(f(\phi)-\frac{\phi}{3}\,\frac{\partial
    f}{\partial\phi}\right)^2 \\[5mm]
        & \displaystyle
        + \frac{16\pi}{3}\,G\,\phi^2\,\left(\frac{\partial
    f}{\partial\phi}\right)^2,
    \end{array}
\end{equation}
which involves the scalar-field potential without supergravity
\begin{equation}\label{m3}
    V(\phi)=\left|\frac{\partial f}{\partial\phi}\right|^2,
\end{equation}
in terms of superpotential $f$.

Putting the bare superpotential equal to
\begin{equation}\label{bare1}
    f_{\mathrm{bare}}=\frac{m_{\mathrm{bare}}}{2\sqrt{2}}\,\phi^2,
\end{equation}
we get the quadric bare potential
\begin{equation}\label{bare2}
    V_{\mathrm{bare}}(\phi)=\frac{m_{\mathrm{bare}}^2}{2}\,\phi^2,
\end{equation}
while the complete bare potential gains the quartic term in accordance with
(\ref{mg2}),
\begin{equation}\label{bare3}
    U_{\mathrm{bare}}(\phi)=\frac{m_{\mathrm{bare}}^2}{2}\,\phi^2+
    \frac{\lambda_{\mathrm{bare}}}{4}\,\phi^4,
\end{equation}
where
\begin{equation}\label{bare4}
    \lambda_{\mathrm{bare}}=\frac{28\pi}{3}\,G\,m_{\mathrm{bare}}^2.
\end{equation}
Therefore, to the linear order in $G$, the supergravity modifies
the quadric bare potential by the quartic term with the constant
\begin{equation}\label{bare5}
    \lambda_{\mathrm{bare}}\sim
    \left(\frac{m_{\mathrm{bare}}}{m_{\mathtt{Pl}}}\right)^2.
\end{equation}
Sure, the bare values run in accordance with both the
renormalization group and redefinitions in the effective
action\footnote{Moreover, the bare mass squared can change its
sing, that can lead to the appearance of local minima in the
potential.}. Thus, they can depend on the field value at low
energies, at least, i.e. when $\phi$ is close to zero.

We accept the nonperturbative or low-energy term\footnote{The
notion of ``low energy'' means the region of potential values
close to zero in comparison with its values during inflation,
say.} of superpotential by setting
\begin{equation}\label{m2}
    f^2_{_\mathrm{LE}}(\phi)=\frac{1}{24\pi\,G}\,\hat \rho(\mathscr{M}),
\end{equation}

Consider the ansatz
\begin{equation}\label{m7}
    \left(\frac{\mathscr{M}}{\mu_{\mbox{\footnotesize\textsc{x}}}}\right)^3=
    1-\left(1-\exp\left\{-\frac{\phi^2}{\widetilde
    m^2}\,[1+\mathcal{C}(\phi)]\right\}\right)^\nu,
\end{equation}
where $\widetilde m$ introduces the mass parameter, while
$\mathcal{C}(\phi)$ is a polynomial function, describing
corrections to the quadratic dependence of the exponent argument
versus the filed.

Then, at $\phi\to 0$ we get $\mathscr{M}\to
\mu_{\mbox{\footnotesize\textsc{x}}}$, and the vacuum density of
energy nullifies, so that at $\nu=3$ the superpotential behaves
like
$$
    f_{_\mathrm{LE}}\sim m_{\mathtt{Pl}}\mu_{\mbox{\footnotesize\textsc{x}}}^2
    \cdot\sqrt{1-\frac{\mathscr{M}}
    {\mu_{\mbox{\footnotesize\textsc{x}}}}}\sim
    \frac{m_{\mathtt{Pl}}\mu_{\mbox{\footnotesize\textsc{x}}}^2}{\widetilde m^3}\,
    \phi^3,
$$
and
$$
    \frac{\partial f_{_\mathrm{LE}}}{\partial\phi}\sim
    \frac{m_{\mathtt{Pl}}\mu_{\mbox{\footnotesize\textsc{x}}}^2}{\widetilde m^3}\,
    \phi^2,
$$
which gives the low-energy correction to the bare quartic
potential
\begin{equation}\label{b-corr}
    V_{_\mathrm{LE}}(\phi)=\frac{\lambda_{_\mathrm{LE}}}{4}\,\phi^4
\end{equation}
with
\begin{equation}\label{b-corr2}
    \lambda_{_\mathrm{LE}}\sim
    \frac{m_{\mathtt{Pl}}^2\mu_{\mbox{\footnotesize\textsc{x}}}^4}{\widetilde
    m^6}.
\end{equation}
Setting
\begin{equation}\label{bare6}
    \lambda_{_\mathrm{LE}}\sim\lambda_{\mathrm{bare}},\quad
    \mbox{and}\quad \widetilde m\sim m_{\mathrm{bare}},
\end{equation}
we find
\begin{equation}\label{bare7}
    \lambda_{_\mathrm{LE}}\sim\frac{\mu_{\mbox{\footnotesize\textsc{x}}}}
    {m_{\mathtt{Pl}}},\qquad
    \widetilde
    m\sim\sqrt{\mu_{\mbox{\footnotesize\textsc{x}}}m_{\mathtt{Pl}}}.
\end{equation}

We consider the situation with thin domain walls corresponding to
the gauge-mediated breaking down SUSY. It suggests that the
correction function $\mathcal{C}(\phi)$ could look as the
expansion in inverse
$\phi_g\sim\mu_{\mbox{\footnotesize\textsc{x}}}$ determined by a
strong-field interaction in the gauge sector, so that to the
leading order one could expect
\begin{equation}\label{mg1}
    \mathcal{C}(\phi)\mapsto \frac{\phi^2}{\phi_g^2}.
\end{equation}
Hence, at $\phi^2\gg\phi_\star^2\sim\widetilde m\phi_g\sim
\widetilde m\mu_{\mbox{\footnotesize\textsc{x}}}$ we arrive to
$\mathscr{M}\to 0$, which gives $U_{_\mathrm{LE}}(\phi)\to
-\rho_{\mbox{\footnotesize\textsc{x}}}$ and
$V_{_\mathrm{LE}}(\phi)\to 0$.

We suppose that the actual superpotential is well approximated by
its low-energy term at $\phi^2<\phi_\star^2$, while at $\phi^2\gg
\phi_\star^2$ the superpotential tends to the bare form. It means
the followings:

\begin{itemize}
    \item The first condition in (\ref{bare6}) should naturally take
    place.
    \item The bare mass in the quadratic term of potential is
    substituted by its running value $m_{\mathrm{bare}}\mapsto
    m(\phi)$, which tends to zero at $\phi^2<\phi_\star^2$, so
    that, at least,
    \begin{equation}\label{running}
    m^2(\phi_\star)\,\phi_\star^2\ll \lambda_{_\mathrm{LE}}\,
    \phi_\star^4,
    \end{equation}
    equivalent to\footnote{Note, that the simplest tadpole diagram due
    to the bare quartic interaction generates the mass determined by
    $m^4\sim\lambda_{_\mathrm{LE}}^2\,\mu_{\mbox{\footnotesize\textsc{x}}}^4$,
    if one puts the cut off about $\mu_{\mbox{\footnotesize\textsc{x}}}$.
    Such the mass will run to the bare value, when the cut off tends to
    the Planck scale.}
    \begin{equation}\label{running2}
    m^4(\phi_\star)\ll \frac{\mu_{\mbox{\footnotesize\textsc{x}}}^5}{
    m_{\mathtt{Pl}}}\sim \lambda_{_\mathrm{LE}}\,\mu_{\mbox{\footnotesize\textsc{x}}}^4,
\end{equation}
    that means
    $m(\phi_\star)\ll\mu_{\mbox{\footnotesize\textsc{x}}}$.
\end{itemize}

Then, the actual potential
$$
    V_{\rm act.}\approx V_{_\mathrm{LE}}(\phi)+\frac{\lambda_{\rm
    bare}}{4}\,\phi^4
$$
acquires a positively-valued extremum at $\phi^2_\star\sim
\widetilde m\mu_{\mbox{\footnotesize\textsc{x}}}$, that means the
breaking down SUSY, while
$$
    U_{\rm act.}\approx U_{_\mathrm{LE}}(\phi)+
    \frac{\lambda_{\rm
    bare}}{4}\,\phi^4
$$
takes a negative value at the extremum,
that guarantees its AdS-position. 

\begin{figure}[th]
  \setlength{\unitlength}{1.0mm}
  \begin{center}
  \hskip13pt
  \begin{picture}(83,42)
  \put(77,11){$\phi$}
  \put(56,11){$\phi_\star$}
  \put(41,35){$U_{\rm act.}(\phi)$}
  \put(41,5){$-\rho_{\mbox{\footnotesize\textsc{x}}}$}
  \put(41,26){$\sqrt{m_{\mathtt{Pl}}^3\mu_{\mbox{\footnotesize\textsc{x}}}^5}$}
  \put(0,0){\includegraphics[width=80\unitlength]{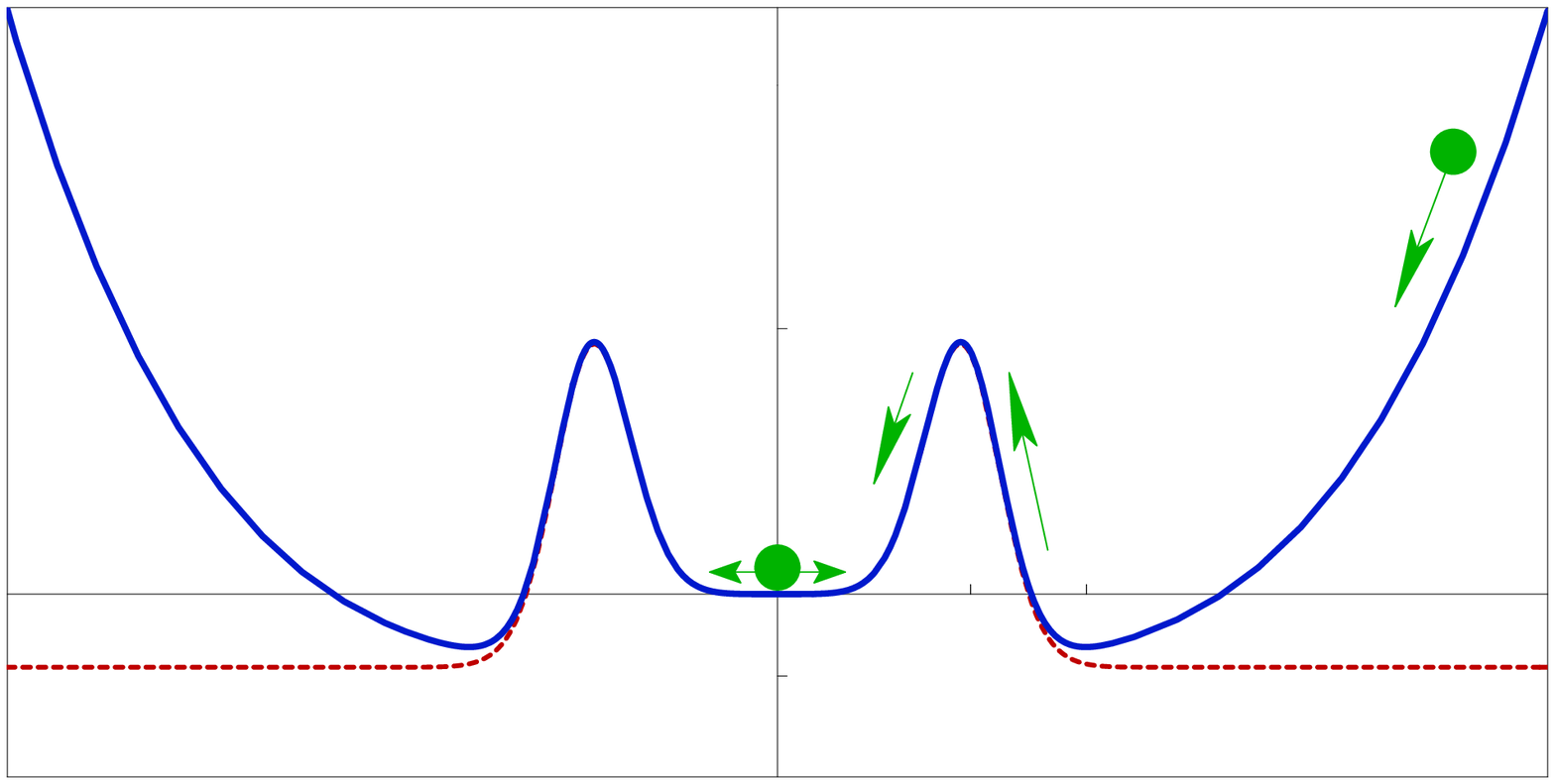}}
  \end{picture}
  \end{center}
  \caption{The actual potential $U_{\rm act.}(\phi)$ with account
  of quartic bare contribution 
  (solid line), and the low-energy term
  $U_{_\mathrm{LE}}(\phi)$ alone (dashed line). The ball and arrows symbolize the field
  evolution to the flat extremum during the inflation and after it
  (see Section \ref{S-IV}).}\label{fall}
\end{figure}

The domain-wall thickness is of the order of
\begin{equation}\label{2l-4}
    \delta r\sim\frac{(\delta\phi)^2}{m_{\mathtt{Pl}}\,
    \mu_{\mbox{\footnotesize\textsc{x}}}^2},
\end{equation}
where $\delta\phi$ is the field increment between the fields
corresponding to the flat and AdS-states, i.e.
$\delta\phi\sim\phi_\star$, hence,
\begin{equation}\label{2l-4a}
    \delta r\sim\frac{\widetilde m}{m_{\mathtt{Pl}}\,
    \mu_{\mbox{\footnotesize\textsc{x}}}}.
\end{equation}
The second condition of (\ref{bare7}) gives
\begin{equation}\label{cond2}
    \delta r\sim\frac{1}{\widetilde m},
\end{equation}
that means that the thickness of domain wall is fixed by the bare
mass of scalar field in the theory.
Then, at $\mu_{\mbox{\footnotesize\textsc{x}}}\sim 10^4$ GeV we
numerically get
\begin{equation}\label{num-cond}
    \lambda\sim 10^{-14},\qquad
    \widetilde m\sim 10^{12}\mbox{ GeV.}
\end{equation}
Fig. \ref{fall} represents the qualitative behavior of actual
potential $U_{\rm act.}(\phi)$ in comparison with the case, when
the bare potential is set to zero.

The local peak of actual potential corresponds to
\begin{equation}\label{peak}
    U_0\sim\frac{f_{_\mathrm{LE}}^2}{\phi_\star^2}\sim
    \sqrt{m_{\mathtt{Pl}}^3\mu_{\mbox{\footnotesize\textsc{x}}}^5}.
\end{equation}

Note, that without the supergravity correction linear in $G$, the
potential remains positive.

\section{The vacuum structure\label{S-III}}

The offered ansatz for the potential gives two AdS-vacuum states
$|\Phi_{\mbox{\footnotesize\textsc{x}}}^\pm\rangle$ positioned in
the extremal points connected by the $\phi\leftrightarrow-\phi$
symmetry. These states possess equal values of energy density
$\rho=-\rho_{\mbox{\footnotesize\textsc{x}}}^{\pm}$. In general,
this condition can be perturbed under constraint of
$\rho_{\mbox{\footnotesize\textsc{x}}}^+\sim\rho_{\mbox{\footnotesize\textsc{x}}}^-$,
which does not essentially break the vacuum texture. The single
flat vacuum-state $|\Phi_{\mbox{\footnotesize\textsc{s}}}\rangle$
is positioned at $\phi=0$ with exact SUSY and $\rho=0$.

The domain-wall fluctuations cause the mixing of such initial
vacuum-states. Indeed, the bubbles of both AdS-vacua can appear in
the flat state. Then, the mixing matrix of vacuum-states takes the
form
\begin{equation}\label{vac1}
    \mathscr{H}=\left(%
    \hskip-2pt
\begin{array}{ccl}
  -\rho_{\mbox{\footnotesize\textsc{x}}}^+ & 0 & \hskip5pt\widetilde
  \rho_+\\[2mm]
  0 & -\rho_{\mbox{\footnotesize\textsc{x}}}^- & \hskip5pt\widetilde
  \rho_-
  \\[2mm]
  \hskip6pt\widetilde \rho_+ & \hskip6pt\widetilde\rho_- & \hskip5pt0 \\
\end{array}%
\hskip-2pt\right),
\end{equation}
where the mixing elements can be taken positive, and by construction
$\widetilde\rho_\pm\ll\rho_{\mbox{\footnotesize\textsc{x}}}^\pm$,
that represents the seesaw mechanism usually applied in the
phenomenology of quarks \cite{Fritzsch}.

Let us put
$\rho_{\mbox{\footnotesize\textsc{x}}}^+=\rho_{\mbox{\footnotesize\textsc{x}}}$
and
$\rho_{\mbox{\footnotesize\textsc{x}}}^-=\rho_{\mbox{\footnotesize\textsc{x}}}(1+u)$
at $u\ll 1$. Therefore, we can separate the leading approximation
suggesting
\begin{equation}\label{vac2}
    \mathscr{H}_0=\left(%
    \hskip-2pt
\begin{array}{ccl}
  -\rho_{\mbox{\footnotesize\textsc{x}}} & 0 & \hskip5pt\widetilde
  \rho_+\\[2mm]
  0 & -\rho_{\mbox{\footnotesize\textsc{x}}} & \hskip5pt\widetilde
  \rho_-
  \\[2mm]
  \hskip6pt\widetilde \rho_+ & \hskip6pt\widetilde\rho_- & \hskip5pt0 \\
\end{array}%
\hskip-2pt
\right),
\end{equation}
and the correction in the form
\begin{equation}\label{vac3}
    \mathscr{V}=-u\cdot \rho_{\mbox{\footnotesize\textsc{x}}}\left(%
\begin{array}{ccl}
  0 & 0 & 0\\
  0 & 1 & 0
  \\
  0 & 0 & 0 \\
\end{array}%
\right),
\end{equation}
so that $\mathscr{H}=\mathscr{H}_0+\mathscr{V}$. Then, the
eigenstate problem can be solved perturbatively in $u\to 0$.

The eigensystem of $\mathscr{H}_0$ is easily found by the
transformation $\mathscr{H}_0\mapsto
\mathscr{H}_0'=\mathscr{U}_0^\dagger\mathscr{H}_0 \mathscr{U}_0$
at
\begin{equation}\label{vac4}
    \mathscr{U}_0=\left(%
    \hskip-5pt
\begin{array}{ccc}
  \hskip7pt\cos\varphi & \sin\varphi & 0 \\[1mm]
  -\sin\varphi & \cos\varphi & 0 \\[1mm]
  \hskip5pt0 & 0 & 1 \\
\end{array}%
\right),
\end{equation}
yielding
\begin{equation}\label{vac5}
    \mathscr{H}_0'=
    \left(%
    \hskip-3pt
\begin{array}{ccc}
  -\rho_{\mbox{\footnotesize\textsc{x}}} & \hskip5pt0 & \hskip5pt0 \\[1.2mm]
  \hskip5pt0 & -\rho_{\mbox{\footnotesize\textsc{x}}} & \hskip5pt\widetilde\rho
  \\[1.2mm]
  \hskip5pt0 & \hskip5pt\widetilde\rho & \hskip5pt0 \\
\end{array}%
\right),
\end{equation}
with
\begin{equation}\label{vac6a}
    \widetilde\rho=\sqrt{\widetilde\rho_-^2+\widetilde\rho_+^2},
\end{equation}
if
\begin{equation}\label{vac6}
    \tan\varphi=\frac{\widetilde\rho_+}{\widetilde\rho_-}.
\end{equation}
Matrix $\mathscr{H}_0'$ in (\ref{vac5}) has got the isolated
$2\raisebox{1pt}{$\scriptstyle\times$} 2$-block, which can be
further transformed to the diagonal form by matrix
\begin{equation}\label{vac7}
    \mathscr{U}=
    \left(%
\begin{array}{ccc}
  1 & \hskip3pt0 & 0 \\[1mm]
  0 & \hskip7pt\cos\theta & \sin\theta \\[1mm]
  0 & -\sin\theta & \cos\theta  \\
\end{array}%
\right),
\end{equation}
at
\begin{equation}\label{vac7a}
    \tan2\theta=\frac{2\widetilde\rho}
    {\rho_{\mbox{\footnotesize\textsc{x}}}}\ll 1.
\end{equation}
Then,
\begin{equation}\label{vac7b}
    \mathscr{U}^\dagger\mathscr{H}_0'\mathscr{U}=\mbox{diag}\left[
    -\rho_{\mbox{\footnotesize\textsc{x}}},\rho_{\rm
    AdS},\rho_{\rm dS}\right],
\end{equation}
where
\begin{equation}\label{vac8}
    \begin{array}{rl}
    \rho_{\rm
    AdS}\hskip-3pt&=-\rho_{\mbox{\footnotesize\textsc{x}}}\cos^2\theta-
    \widetilde\rho\sin2\theta,\\[2mm]
    \rho_{\rm
    dS}\hskip-3pt&=-\rho_{\mbox{\footnotesize\textsc{x}}}\sin^2\theta+
    \widetilde\rho\sin2\theta,\\[2mm]
    \end{array}
\end{equation}
that can be further expanded in $\theta\to 0$ due to
(\ref{vac7a}),
\begin{equation}\label{vac8a}
    \begin{array}{rl}
    \rho_{\rm
    AdS}\hskip-3pt&\approx-\rho_{\mbox{\footnotesize\textsc{x}}}-
    \displaystyle
    \frac{\widetilde\rho^2}
    {\rho_{\mbox{\footnotesize\textsc{x}}}},\\[4mm]
    \rho_{\rm
    dS}\hskip-3pt&\approx\displaystyle
    \frac{\widetilde\rho^2}
    {\rho_{\mbox{\footnotesize\textsc{x}}}}=\frac{\widetilde\rho_-^2+\widetilde\rho^2_+}
    {\rho_{\mbox{\footnotesize\textsc{x}}}}.
    \end{array}
\end{equation}
The eigenstates are determined by the product of rotations
$\mathscr{U}_0 \mathscr{U}$ as follows:
\begin{equation}\label{vac9}
    \begin{array}{lcl}
    \hskip-9pt
      |\Phi_{\rm AdS}'\rangle&\hskip-5pt=&\hskip-4pt
      \cos\varphi|\Phi_{\mbox{\footnotesize\textsc{x}}}^+\rangle-
      \sin\varphi|\Phi_{\mbox{\footnotesize\textsc{x}}}^-\rangle,\\[2.5mm]
      \hskip-9pt
      |\Phi_{\rm AdS}\rangle&\hskip-5pt=&\hskip-4pt
      \cos\theta\left\{\sin\varphi|\Phi_{\mbox{\footnotesize\textsc{x}}}^+\rangle+
      \cos\varphi|\Phi_{\mbox{\footnotesize\textsc{x}}}^-\rangle\right\}-\sin\theta
      |\Phi_{\mbox{\footnotesize\textsc{s}}}\rangle,\\[2.5mm]
      \hskip-9pt
      |\Phi_{\rm dS}\rangle&\hskip-5pt=&\hskip-4pt
      \sin\theta\left\{\sin\varphi|\Phi_{\mbox{\footnotesize\textsc{x}}}^+\rangle+
      \cos\varphi|\Phi_{\mbox{\footnotesize\textsc{x}}}^-\rangle\right\}+\cos\theta
      |\Phi_{\mbox{\footnotesize\textsc{s}}}\rangle.\\[-4mm]
      &&
    \end{array}
\end{equation}
In the simplest realization of potential with the incorporation of
$\phi\leftrightarrow-\phi$ symmetry, we get $\widetilde
\rho_-=\widetilde\rho_+$ in $\mathscr{H}_0$, so that
$\varphi=\pi/4$. Remember, that for thin domain walls we evaluate
the mixing elements in \cite{KT1,KT2} by
$$
    \widetilde
    \rho\sim\frac{\mu_{\mbox{\footnotesize\textsc{x}}}^6}{m_\mathtt{Pl}^2},
$$
so that $\mu_{\mbox{\footnotesize\textsc{x}}}\sim 10^4$ GeV is
consistent with the observed scale of cosmological constant. Such
the kind of relation between the scales of SUSY breaking down and
cosmological constant was derived by T.Banks \cite{Banks-I} in
other way of argumentations.

The corrections to the energy densities linear in $u$ are
straightforwardly determined by appropriate diagonal matrix
elements of perturbation $\mathscr{V}$, so that
\begin{equation}\label{vac10}
    \begin{array}{rcl}
    \hskip-9pt
      \delta\rho^\prime_{\rm AdS} & \hskip-5pt= &\hskip-3pt
      -u\cdot\sin^2\varphi\cdot\rho_{\mbox{\footnotesize\textsc{x}}},
      \\[5mm]
      \delta\rho_{\rm AdS} & \hskip-5pt= &\hskip-3pt
      -u\cdot\cos^2\varphi\,\cos^2\theta\cdot\rho_{\mbox{\footnotesize\textsc{x}}}
      \approx -u\cdot\cos^2\varphi\cdot\rho_{\mbox{\footnotesize\textsc{x}}},
      \\[3mm]\displaystyle
      \delta\rho_{\rm dS} & \hskip-5pt= & \hskip-3pt\displaystyle
      -u\cdot\cos^2\varphi\,\sin^2\theta\cdot\rho_{\mbox{\footnotesize\textsc{x}}}
      \approx -u\cdot\cos^2\varphi\,\frac{\widetilde\rho^2}
      {\rho_{\mbox{\footnotesize\textsc{x}}}}.\\[-6mm]
      &&
    \end{array}
\end{equation}
Non-diagonal matrix elements of perturbation result in the mixing
of states defined in (\ref{vac9}). These elements take hermitian
values,
\begin{equation}\label{vac11}
    \begin{array}{rcl}
    \hskip-9pt
    \langle \Phi_{\rm AdS}|\mathscr{V}|\Phi^\prime_{\rm
    AdS}\rangle& \hskip-5pt= & \hskip-3pt\displaystyle
    \frac{u}{2}\,\rho_{\mbox{\footnotesize\textsc{x}}}\cdot\sin2
    \varphi\cos\theta,\\[4mm]
    \langle \Phi_{\rm dS}|\mathscr{V}|\Phi^\prime_{\rm
    AdS}\rangle& \hskip-5pt= & \hskip-3pt\displaystyle
    \frac{u}{2}\,\rho_{\mbox{\footnotesize\textsc{x}}}\cdot
    \sin2\varphi\,
    \sin\theta,\\[4mm]-
    \langle \Phi_{\rm dS}|\mathscr{V}|\Phi_{\rm
    AdS}\rangle& \hskip-5pt= & \hskip-3pt\displaystyle
    \frac{u}{2}\,\rho_{\mbox{\footnotesize\textsc{x}}}\cdot
    \cos^2\varphi\,
    \sin2\theta,
    \\[-5mm]
      &&
    \end{array}
\end{equation}
so that at $\theta\ll 1$ we approximately get
\begin{equation}\label{vac9d}
    \begin{array}{rcl}
    \hskip-9pt
      \delta|\Phi_{\rm
      AdS}'\rangle&\hskip-5pt\approx&\hskip-4pt\displaystyle
      \frac{u\,\sin2\varphi}{2\sin^2\theta}\left\{|\Phi_{\rm
      AdS}\rangle-\sin^3\theta
      |\Phi_{\rm dS}\rangle\right\}\hskip-2pt,\\[4.5mm]
      \hskip-9pt
      \delta|\Phi_{\rm
      AdS}\rangle&\hskip-5pt\approx&\hskip-4pt\displaystyle
      \frac{u\,\sin2\varphi}{2\sin^2\theta}\left\{-|\Phi_{\rm AdS}^\prime\rangle+
      \cot\varphi\, \sin^3\theta|\Phi_{\rm dS}\rangle\right\}\hskip-2pt,\\[4.5mm]
      \hskip-9pt
      \delta|\Phi_{\rm
      dS}\rangle&\hskip-5pt\approx&\hskip-4pt\displaystyle u
      \sin\theta\cos\varphi\left\{\sin\varphi|\Phi_{\rm
      AdS}^\prime\rangle-
      \cos\varphi|\Phi_{\rm AdS}\rangle\right\}\hskip-2pt,\\[-0mm]
      &&
    \end{array}
\end{equation}
wherein one could further expand in $\theta$, replacing
$\sin\theta$ by $\theta$ itself. Moreover, we can give the above
corrections to the states in terms of initial basis
$\{|\Phi_{\mbox{\footnotesize\textsc{x}}}^+\rangle,
|\Phi_{\mbox{\footnotesize\textsc{x}}}^-\rangle,
|\Phi_{\mbox{\footnotesize\textsc{s}}}\rangle\}$, so that
$$
    \hspace*{3mm}\begin{array}{rcl}
    \hskip-9pt
      \delta|\Phi_{\rm
      AdS}'\rangle&\hskip-5pt\approx&\hskip-4pt\displaystyle
      \frac{u\,\sin2\varphi}{2\theta^2}\left\{\sin\varphi|
      \Phi_{\mbox{\footnotesize\textsc{x}}}^+\rangle+\cos\varphi
      |\Phi_{\rm dS}^-\rangle-\theta|\Phi_{\mbox{\footnotesize\textsc{s}}}\rangle
      \right
      \}\hskip-2pt,\\[4.5mm]
      \hskip-9pt
      \delta|\Phi_{\rm
      AdS}\rangle&\hskip-5pt\approx&\hskip-4pt\displaystyle
      \frac{u\,\sin2\varphi}{2\theta^2}\left\{-\cos\varphi|
      \Phi_{\mbox{\footnotesize\textsc{x}}}^+\rangle+\sin\varphi
      |\Phi_{\mbox{\footnotesize\textsc{x}}}^-\rangle\right.\\[2mm]
      &&\hskip2.92cm\displaystyle\left.
      -\theta^3\cot\varphi|\Phi_{\mbox{\footnotesize\textsc{s}}}\rangle
      \right
      \}\hskip-2pt,\\[4.5mm]
      \hskip-9pt
      \delta|\Phi_{\rm
      dS}\rangle&\hskip-5pt\approx&\hskip-4pt\displaystyle u\theta\,\cos\varphi
      \hskip-1.8pt\left\{\theta^2\hskip-0.8pt\sin2\varphi|
      \Phi_{\mbox{\footnotesize\textsc{x}}}^+\rangle-
      |\Phi_{\mbox{\footnotesize\textsc{x}}}^-\rangle+
      \theta\cos\varphi|\Phi_{\mbox{\footnotesize\textsc{s}}}\rangle
      \hskip-1pt\right
      \}\hskip-2.6pt.\\[0mm]
      &&
    \end{array}\nonumber
$$

Therefore, the mixing remains under the control of perturbative
theory, if
\begin{equation}\label{vac12}
    u\ll\sin^2\theta,
\end{equation}
i.e., when the splitting between
$\rho_{\mbox{\footnotesize\textsc{x}}}^+$ and
$\rho_{\mbox{\footnotesize\textsc{x}}}^-$ is much less than
$\rho_{\rm dS}$. Otherwise, the texture of (\ref{vac9}) is
essentially changed. Nevertheless, we can generically expect that
the stationary AdS-levels include the mixtures of initial
$|\Phi_{\mbox{\footnotesize\textsc{x}}}^\pm\rangle$-states with
amplitudes of the order of unit, while the contribution of flat
state $|\Phi_{\mbox{\footnotesize\textsc{s}}}\rangle$ is
suppressed. The dS-level is dominantly represented by the flat
state, while the SUSY breaking amplitudes give a suppressed
admixture.

Some other approaches to the cosmological constant problem in the
framework of seesaw mechanism can be found in
\cite{Enqvist:2007tb,Smolyakov}.

\section{Incorporating the inflation\label{S-IV}}

The standard scenario of inflation with the single scalar field
$\phi$ possessing the quadratic and quartic self-couplings
suggests that the inflaton takes values about the Planck scale. As
we have mentioned above, at such the field values the running mass
can be close to its bare value, that leads to approximate equality
of quadratic and quartic terms of potential. Then, we get well
elaborated picture of inflation consistent with observed features
of matter inhomogeneity in the Universe: the magnitude of
fluctuations of matter density, spectral index of scalar
fluctuations and fraction of tensor perturbations, if one takes
the couplings in agreement with (\ref{num-cond}). Thus, the
parameters of inflaton potential are inherently related with the
scale of SUSY breaking down.

However, a numerical analysis should be performed with more caution,
since the relation between the scales in (\ref{bare4}) due to the
simplest connection of bare quadric potential with the quartic
coupling suggests a too steep growth of quartic term, which starts
to dominate during the inflation. But the quartic-potential
inflation is almost inconsistent with the data obtained by WMAP
\cite{WMAP5-1,WMAP5-2,WMAP7} and experiments on baryonic acoustic
oscillations and spacial distribution of galaxies (BAO) \cite{BAO}
as well as on the supernovae Ia (SN)
\cite{Riess:2004nr,Riess:2006fw,Astier:2005qq,WoodVasey:2007jb}, and
it can be  marginally accepted, only. Then, one should consider a
mechanism for the dominance of quadratic term alone, that is
still consistent with the data.

Let us consider the problem in more detail. So, at Planckian
values of scalar field we get the potential in the form
\begin{equation}\label{inf1}
    V=\frac{m^2}{2}\,\phi^2+\frac{\lambda}{4}\,\phi^4,
\end{equation}
with positive mass squared $m^2>0$. In the homogeneous
Friedmann--Robertson--Walker metric
\begin{equation}\label{inf2}
    \mathrm{d}s^2=\mathrm{d}t^2-a^2(t)\mathrm{d}\boldsymbol r^2
\end{equation}
with the scale factor of expansion $a(t)$, the standard evolution
equations with respect to time read off as followings:

\begin{itemize}
    \item The field runs according to
\begin{equation}\label{inf3}
    \ddot \phi+3 H\,\dot\phi+m^2\phi+\lambda\phi^3=0,
\end{equation}
where the over-dot denotes the derivative with respect to time
$\mathrm{d}\triangledown/\mathrm{d}t\equiv \dot\triangledown$, and
the Hubble rate is defined by $H=\dot a/a$.
    \item The Friedmann equation determines the rate of expansion
\begin{equation}\label{inf4}
    H^2=\frac{\kappa^2}{3}\left\{\frac{1}{2}\,\dot\phi^2
    +\frac{m^2}{2}{\,\phi^2+
    \frac{\lambda}{4}\,\phi^4}\right\},
\end{equation}
with $\kappa^2=8\pi G$.
    \item The acceleration of expansion is given by
\begin{equation}\label{inf5}
    \dot H=-\frac{1}{2}\,\kappa^2\dot\phi^2.
\end{equation}
\end{itemize}

During the inflation the kinetic energy is suppressed with respect
to the potential term, hence, the Hubble rate slowly changes in
accordance with (\ref{inf5}), so it is almost a constant.

It is spectacular that the homogeneous evolution demonstrates the
behavior of parametric attractor: the kinetic and potential energies of
inflaton rapidly tend to definite critical values independent of
initial data, while the critical points gain a driftage with the
slowly changing Hubble rate\footnote{The dependence of inflation
on initial data were originally studied in
\cite{BGKhZ,PW,KLS,laM-P,KBranden}.}. In order to show this fact,
we follow the method developed in \cite{Mexicans,KT3} and introduce
appropriate scaling variables
\begin{equation}\label{inf6}
    x=\frac{\kappa}{\sqrt{6}}\,\frac{\dot\phi}{H},\qquad
    y^2=\frac{\kappa^2}{12 H^2}\;\big(2\,{m^2}\,\phi^2+
    \lambda\,\phi^4\big),
\end{equation}
as well as the control parameter of driftage
\begin{equation}\label{inf7}
    z^4=\frac{3\lambda}{\kappa^2H^2}.
\end{equation}
Defining the amount of e-folding to the end of inflation by $N=\ln
a_\mathrm{end}-\ln a$ and denoting the derivative with respect to
$N$ by prime $\mathrm{d}\diamond/\mathrm{d}N\equiv
\diamond^\prime$, we get the evolution equations of autonomous
system with the parameter $z$
\begin{equation}\label{inf77}
    \begin{array}{rcl}
      x^\prime & \hskip-3pt = & \hskip-3pt -3x^3+3x+2 z\,
      \xi(y,z)\,\zeta(y,z),  \\[2mm]
      y\,y^\prime & \hskip-3pt = & \hskip-3pt -3x^2y^2-2x z\,
      \xi(y,z)\,\zeta(y,z),  \\[2mm]
    \end{array}
\end{equation}
where
\begin{equation}\label{inf7a}
    \begin{array}{rcl}
      \xi^2(y,z) & \hskip-3pt = & \hskip-3pt y^2+u^4z^4,  \\[2mm]
      \zeta^2(y,z) & \hskip-3pt = & \xi(y,z)-u^2z^2,  \\[2mm]
    \end{array}
\end{equation}
while
\begin{equation}\label{inf7b}
    z^\prime=-\frac{3}{2}\,x^2z,
\end{equation}
and $u$ is the constant parameter defined by
\begin{equation}\label{inf7c}
    u^2=\frac{\kappa^2m^2}{6\lambda}.
\end{equation}
Hence, the field takes the values
$$
    \phi=\pm\frac{m\zeta(y,z)}{uz\sqrt{\lambda}}.
$$
The Friedmann condition of (\ref{inf4}) is transformed into
\begin{equation}\label{inf4a}
    x^2+y^2=1,
\end{equation}
that is conserved by the dynamical system of
(\ref{inf77})--(\ref{inf7c}), of course.

The acceleration takes place at $\ddot a>0$, that gives
$$
    \frac{\ddot a}{a}=\dot H+H^2>0,
$$
equivalent to $-\dot H/H^2<1$ yielding the condition of inflation
end
\begin{equation}\label{inf8}
    x^2<\frac{1}{3}.
\end{equation}
The nonzero critical points $\{x_c,y_c\}$ put
$x^\prime=y^\prime=0$, and they are positioned on the Friedmann
circle of (\ref{inf4a}) and related by
\begin{equation}\label{inf9}
    3x_cy_c^2=-2z\,\xi(y_c,z)\,\zeta(y_c,z).
\end{equation}
Linear perturbations $x=x_c+\bar x$ and $y=y_c+\bar y$ in
(\ref{inf77}) give
$$
   \left(%
\begin{array}{c}
  \bar x^\prime \\
  \bar y^\prime \\
\end{array}%
\right)=
\left(%
\begin{array}{cc}
  3-9x_c^2 & \displaystyle-\frac{2z^2}{3x_cy_c}(3\xi_c-2u^2z^2)
  \\[4mm]
  -3x_cy_c & \displaystyle\frac{2z^2}{3x_cy_c}(3\xi_c-2u^2z^2)-6x_c^2 \\
\end{array}%
\right)
\left(%
\begin{array}{c}
  \bar x \\
  \bar y \\
\end{array}%
\right)
$$
so that under the constraint of (\ref{inf4a}) resulting in the
relation $x_c\bar x+y_c\bar y=0$, we find
\begin{equation}\label{inf10}
    \bar x^\prime=\gamma_c\cdot \bar x,
\end{equation}
at
\begin{equation}\label{inf11}
    \gamma_c=3-9x_c^2-\frac{x_c
    z}{\xi_c\zeta_c}\,(2\zeta_c^2+\xi_c),
\end{equation}
where we put $\xi_c=\xi(y_c,z)$ and $\zeta_c=\zeta(y_c,z)$. Then,
perturbations exponentially decline with the expansion as
\begin{equation}\label{inf12}
    \bar x= \bar
    x_\mathrm{tot.}\mathrm{e}^{-\gamma_c(N_\mathrm{tot.}-N)},
\end{equation}
if
\begin{equation}\label{inf13}
    \gamma_c>0,
\end{equation}
yielding
\begin{equation}\label{inf14}
    x_c^2\left(1-\frac{y_c^4}{6}\,\frac{2\zeta_c^2+\xi_c}{\zeta_c^2\xi_c^2}
    \right)<\frac{1}{3},
\end{equation}
consistent with the condition of inflationary expansion
(\ref{inf8}). Therefore, the critical point is stable during the
inflation, and we have got the quasi-atractor at
\begin{equation}\label{attract1}
    x_c^2\ll 1, \quad y_c^2\approx 1,
\end{equation}
because the control parameter has got a slow driftage as
\begin{equation}\label{drift}
    z\approx z_\star\left\{1-\frac{3}{2}x_c^2(N-N_\star)\right\},
\end{equation}
effective at large intervals
$$
    |N-N_\star|\ll\frac{1}{x_c^2}.
$$

\noindent We are interested in two limits:\\

\begin{tabular}{ll}
  $u^2z^2\ll 1$ & quartic term dominance, \\[2mm]
  $u^2z^2\gg 1$ & quadratic term dominance.\\[1mm]
\end{tabular}

\noindent So, at $y\sim1$
\begin{equation}\label{inf15}
    \xi=\left\{
    \begin{array}{cl}
      y, & u^2z^2\ll 1, \\[4mm]
      u^2z^2, & u^2z^2\gg 1, \\
    \end{array}
    \right.
\end{equation}
and
\begin{equation}\label{inf15a}
    \zeta=\left\{
    \begin{array}{cl}
      \sqrt{y}, & u^2z^2\ll 1, \\[4mm] \displaystyle
      \frac{y}{uz\sqrt{2}}, & u^2z^2\gg 1. \\
    \end{array}
    \right.
\end{equation}
Then, the attractor stability takes place at
\begin{equation}\label{inf16}
      \begin{array}{ccl}\displaystyle
      x_c^2<\frac{2}{3}, && u^2z^2\ll 1, \\[4mm]\displaystyle
      x_c^2<\frac{1}{2}, && u^2z^2\gg 1. \\
    \end{array}
\end{equation}

The amount of e-foldings is accurately approximated by
\begin{equation}\label{inf17}
    N\approx -\frac{2}{3}\int\limits^z\frac{\mathrm{d}{z}}{x_c^2
    z},
\end{equation}
with (\ref{inf9}), so that
\begin{equation}\label{inf18}
    N\approx\left\{
      \begin{array}{ccl}\displaystyle
      \frac{3}{4 z^2}, && u^2z^2\ll 1, \\[4mm]\displaystyle
      \frac{3}{4 u^2z^4}, && u^2z^2\gg 1. \\
    \end{array}
    \right.
\end{equation}

The inhomogeneities are approximated by the scalar and tensor
densities of spectra versus the wave-vector at $k=a(t)\,H$ as
followings:
\begin{equation}\label{inh-1}
    \begin{array}{rcl}
      \mathcal{P}_{\mathrm{S}}(k) & \hskip-3pt= & \hskip-3pt\displaystyle
      \left(\frac{H}{2\pi}\right)^2\left(\frac{H}{\dot\phi}\right)^2,\\[5mm]
      \mathcal{P}_{\mathrm{T}}(k) & \hskip-3pt= & \hskip-3pt\displaystyle
      8\kappa^2\left(\frac{H}{2\pi}\right)^2,
    \end{array}
\end{equation}
which can be accurately evaluated in terms of quasi-attractor
dynamics by
\begin{equation}\label{inh-2}
    \begin{array}{rcl}
      \mathcal{P}_{\mathrm{S}}(k) & \hskip-3pt= & \hskip-3pt\displaystyle
      \frac{\lambda}{8\pi^2}\,\frac{1}{z^4 x_\mathrm{c}^2},\\[5mm]
      \mathcal{P}_{\mathrm{T}}(k) & \hskip-3pt= & \hskip-3pt\displaystyle
      \frac{6\lambda}{\pi^2}\,\frac{1}{z^4},
    \end{array}
\end{equation}
while the ratio
\begin{equation}\label{inh-5}
    r=\frac{\mathcal{P}_{\mathrm{T}}}{\mathcal{P}_{\mathrm{S}}}=48\,x_{\mathrm{c}}^2
    \ll 1,
\end{equation}
and it determines the relative contribution of tensor spectrum.

The spectral index of scalar spectrum is defined by
\begin{equation}\label{inh-7}
    n_{\mathrm{S}}-1\equiv \frac{\mathrm{d}\ln
    \mathcal{P}_{\mathrm{S}}}{\mathrm{d}\ln k}.
\end{equation}
It can be calculated under the condition
\begin{equation}\label{inh-7-1}
    \ln\frac{k}{k_{\mathrm{end}}}=-N-2\ln\frac{z}{z_{\mathrm{end}}},
\end{equation}
which gives
$$
    \frac{\mathrm{d}\ln k}{\mathrm{d}N}\approx
    -1,
$$
to the leading order in $1/N$.

Then, following (\ref{inf11}), (\ref{inf15}), (\ref{inf15a}),
(\ref{inf18}), we get the limits,
\begin{equation}\label{res}
    \begin{array}{|c|c|c|c|}
      \hline
      \mbox{limit} & \hskip3mm\mathcal{P}_{\mathrm{S}}(k) \hskip3mm
      & \hskip3mm n_{\mathrm{S}}-1 \hskip3mm& \hskip3mm
      r\hskip3mm\\[1mm]
      \hline
      &&&\\[-3mm]
       u^2z^2\ll 1 & \displaystyle
      \frac{2\lambda}{3\pi^2}\,
      {N^3} & \displaystyle -\frac{3}{N} &
      \displaystyle\frac{16}{N}\\[3mm]
      \hline
      &&&\\[-3mm]
      u^2z^2\gg 1 & \displaystyle
      \frac{\lambda u^2}{\pi^2}\,
      {N^2} & \displaystyle -\frac{2}{N} &
      \displaystyle\frac{8}{N}\\[3mm]
      \hline
    \end{array}
\end{equation}

The data on the correlation in the plain of $\{n_{\mathrm{S}},r\}$
\cite{Weinberg-RMP,WMAP5-1,WMAP5-2,WMAP7,BAO,Riess:2004nr,Riess:2006fw,
Astier:2005qq,WoodVasey:2007jb}
prefer for the case of quadric coupling at $N\approx 60$ \cite{LL},
while the quartic coupling alone is marginally consistent with the
data under $N\approx 80$, which is rather unrealistic \cite{KT3}.

Therefore, we conclude that realistic scenario suggests $u^2z^2\gg
1$, i.e. $u^4z^4\gg1$, which results in
\begin{equation}\label{res2}
    u^2\gg \frac{4}{3}\,N,
\end{equation}
due to (\ref{inf18}). From the relation between the bare values in
(\ref{bare4}), we get
\begin{equation}\label{res3}
    u^2_\mathrm{bare}=\frac{1}{7},
\end{equation}
which is inconsistent with the referable dominance of quadratic
term. However, we have to take into account the running of
potential parameters mentioned above. In this respect one could
put the quartic constant approximately equal to its bare value,
since this value is extremely small, so that one could expect no
significant renormalization of the constant even at large
logarithmic increment of scale. In contrast, one can put
\begin{equation}\label{res4}
    m(m_\mathrm{bare})\sim m_\mathrm{bare},\qquad
    m(m_\mathtt{Pl})\sim K\cdot m_\mathrm{bare},
\end{equation}
at $K\gg1$. Then,
\begin{equation}\label{res5}
    u^2\sim \frac{K^2}{7},
\end{equation}
and the quadratic term dominates, if
\begin{equation}\label{rere}
    K^2\gg 10\cdot N\sim 600,
\end{equation}
or $K\gg 25$, hence, the actual mass of inflaton should be about $10^{13}$ GeV
in agreement with the phenomenological analysis of observed data. 
Such the situation is not in any contradiction with
quite general properties of the potential, as concerns its scales,
since factors of the form $4\pi^2$ could be responsible for the
finite rescaling used above. Thus, the dominance of quadratic term
can be actual.

One could simply require (\ref{res4}) from the form of potential
at $\phi\sim m_\mathtt{Pl}$ by setting $m^2\gg \lambda\,
m_\mathtt{Pl}^2\sim
\mu_{\mbox{\footnotesize\textsc{x}}}m_\mathtt{Pl}\sim
m^2_\mathrm{bare}$, of course. However, the estimate of
(\ref{rere}) can be derived from the analysis presented above,
only.

Another aspect of inflation is related with the potential behavior
in vicinity of $\phi=0$. First, the AdS minimum is not essential
for the inflation, since its scale of energy density
$\rho_{\mbox{\footnotesize\textsc{x}}}\sim\mu_{\mbox{\footnotesize\textsc{x}}}^4$
is essentially less than the energy density to the end of
inflation $\rho_\mathrm{end}\sim \lambda\, m_\mathtt{Pl}^4\sim
\mu_{\mbox{\footnotesize\textsc{x}}}m_\mathtt{Pl}^3$, i.e. there
is the hierarchy $\rho_{\mbox{\footnotesize\textsc{x}}}\ll
\rho_\mathrm{end}$. Second, the potential barrier between the flat
and AdS vacua has the height about
$$
    U_0\sim
    \sqrt{m_{\mathtt{Pl}}^3\mu_{\mbox{\footnotesize\textsc{x}}}^5}\sim
    \frac{\mu_{\mbox{\footnotesize\textsc{x}}}^3}{m_\mathtt{Pl}^3}\,\rho_\mathrm{end}
    \ll \rho_\mathrm{end}.
$$
Therefore, the local peak of potential is also inessential for the
inflation, too. However, both these items can be involved into the
mechanism of reheating.

In this respect, the most essential effect is related with the
potential barrier, since near the peak the effective inflaton-mass
squared is negative, that causes the tachyonic instability
resulting in the preheating mechanism: a rapid decay of the
inflaton to quanta at the moment, when the energy density becomes
comparable with the peak height \cite{tach-preh}. Then, we can
evaluate the preheating temperature $T_\mathrm{preh.}$ by
\begin{equation}\label{rere2}
    T_\mathrm{preh.}^4\sim U_0,\qquad \mbox{if}\quad
    m_\mathrm{eff.}\ll T_\mathrm{preh.},
\end{equation}
where $m_\mathrm{eff.}$ is an effective mass of quanta.
Numerically, (\ref{rere2}) gives
\begin{equation}\label{rere3}
    T_\mathrm{preh.}\sim 10^9\mbox{ GeV.}
\end{equation}
In the model under consideration, the inflaton quanta with respect
to the flat vacuum possess the effective mass equal to zero,
$$
    m_\mathrm{eff.}^\mathrm{flat}=0,
$$
while the quanta with respect to the AdS vacuum have nonzero mass,
that can be roughly evaluated by
$$
    \left\{m_\mathrm{eff.}^\mathrm{AdS}\right\}^2\sim\frac{\partial^2
    U}{\partial\phi^2}
    \sim\frac{U_0}{\phi_\star^2},
    \quad\Rightarrow\quad
    m_\mathrm{eff.}^\mathrm{AdS}\sim m_\mathrm{bare},
$$
though the mass can have got a much less value. Nevertheless, the
decay to massless quanta would be kinematically preferable, that
leads to \textit{the relaxation of inflaton in vicinity of flat
vacuum}, as pictured in Fig. \ref{fall}.

Thus, the scalar field could inflationary evolve and relax in the
flat vacuum after the tachyonic preheating in agreement with
current experimental constraints from the observational data. Such
the evolution could explain why we are living in the vacuum we
have got.

A discussion of some other aspects of inflation in supergravity
can be found in review \cite{inflation}. For instance, the
mechanism with an effectively real inflaton was considered by
Kawasaki, Yamaguchi and Yanagidain in \cite{KYY}, while the
mechanism for the vacuum stabilization was offered by Kachru,
Kallosh, Linde and Trevedi in \cite{KKLT}. Further developments
include also the problem of irreversible vacuum decays that serve
as sinks for the probability flow \cite{sinks}.

\section{Three generations\label{S-V}}

The vacuum structure considered in Section \ref{S-III} suggests
that quantum field vibrations in vicinity of initial vacuum-states
can mix. We can easily investigate the main features of such the
mixing in the simplest case of potential symmetry versus
$\phi\leftrightarrow-\phi$. Then, the mass matrix for fermions
takes the form
\begin{equation}\label{gen1}
    \mathcal{M}=
    \left(%
\begin{array}{ccc}
  m_{\mbox{\footnotesize\textsc{x}}}\hskip2pt &
  \mu_{\mbox{\footnotesize\textsc{a}}} &
  \mu_{\mbox{\footnotesize\textsc{b}}}' \\
  \mu_{\mbox{\footnotesize\textsc{a}}}\hskip3pt &
  m_{\mbox{\footnotesize\textsc{x}}} &
  \mu_{\mbox{\footnotesize\textsc{b}}} \\
  {\mu_{\mbox{\footnotesize\textsc{b}}}'}^* &
  \mu_{\mbox{\footnotesize\textsc{b}}} &
  m_{\mbox{\footnotesize\textsc{s}}} \\
\end{array}%
\right),
\end{equation}
where $m_{\mbox{\footnotesize\textsc{x}}}$ stands for the fermion
mass in the AdS-vacua, when SUSY is broken down, while
$m_{\mbox{\footnotesize\textsc{s}}}$ denotes the fermion mass in
the flat supersymmetric vacuum. Elements
$\mu_{\mbox{\footnotesize\textsc{a,b}}}$ introduce the mixing. In
(\ref{gen1}) all of elements except
$\mu_{\mbox{\footnotesize\textsc{b}}}'$ are real due to the
freedom in the definition of complex phases for the initial
states, while
$|\mu_{\mbox{\footnotesize\textsc{b}}}'|=\mu_{\mbox{\footnotesize\textsc{b}}}$
because of symmetry.

Let us introduce the complex phase $\gamma$ by setting
\begin{equation}\label{gen2}
    \mu_{\mbox{\footnotesize\textsc{b}}}'=\mu_{\mbox{\footnotesize\textsc{b}}}\,
    \mathrm{e}^{\mathrm{i}\gamma}.
\end{equation}
Transforming the matrix to $\mathcal{M}_\mathcal{U}=
\mathcal{U}\cdot \mathcal{M}\cdot \mathcal{U}^\dagger$ with
\begin{equation}\label{gen3}
    \mathcal{U}=
    \left(%
\begin{array}{ccc}
  c_0 & s_0^+ & 0 \\
  -s_0^- & c_0 & 0 \\
  0 & 0 & 1 \\
\end{array}%
\right)
\end{equation}
at
\begin{equation}\label{gen4}
    c_0=-s_0=\frac{1}{\sqrt{2}},\quad
    s_0^+=s_0\,\mathrm{e}^{\mathrm{i}\gamma},\quad
    s_0^-=s_0\,\mathrm{e}^{-\mathrm{i}\gamma},
\end{equation}
we get
\begin{equation}\label{gen5}
    \mathcal{M}_\mathcal{U}=\hskip-2pt
    \left(%
    \hskip-1pt
\begin{array}{ccc}
  m_{\mbox{\footnotesize\textsc{x}}}-\mu_{\mbox{\footnotesize\textsc{a}}}\cos\gamma &
  -\mathrm{i}\mu_{\mbox{\footnotesize\textsc{a}}}\sin\gamma\,
  \mathrm{e}^{\mathrm{i}\gamma} & 0 \\[2mm]
  \mathrm{i}\mu_{\mbox{\footnotesize\textsc{a}}}\sin\gamma\,
  \mathrm{e}^{-\mathrm{i}\gamma} &
  m_{\mbox{\footnotesize\textsc{x}}}+\mu_{\mbox{\footnotesize\textsc{a}}}\cos\gamma &
  \sqrt{2} \mu_{\mbox{\footnotesize\textsc{b}}} \\[2mm]
  0 & \sqrt{2}\mu_{\mbox{\footnotesize\textsc{b}}} & m_{\mbox{\footnotesize\textsc{s}}} \\
\end{array}%
\hskip-3pt \right)\hskip-3pt.
\end{equation}
From (\ref{gen5}) we see that the analysis is essentially
simplified at $\gamma=\{0,\pm\pi\}$, when one can neglect effects
caused by violation of combined invariance with respect to the
charge conjugation $\mathbb{C}$ and mirror inversion of space
$\mathbb{P}$, and the matrix takes the symmetric form, so that at
$\gamma=\pm\pi$
\begin{equation}\label{gen6}
    \mathcal{M}_\mathcal{U}^{(0)}=\hskip-1pt
    \left(%
\begin{array}{ccc}
  m_{\mbox{\footnotesize\textsc{x}}}+\mu_{\mbox{\footnotesize\textsc{a}}} &
  0 & 0 \\[2mm]
  0 &
  m_{\mbox{\footnotesize\textsc{x}}}-\mu_{\mbox{\footnotesize\textsc{a}}} &
  \sqrt{2} \mu_{\mbox{\footnotesize\textsc{b}}} \\[2mm]
  0 & \sqrt{2}\mu_{\mbox{\footnotesize\textsc{b}}} & m_{\mbox{\footnotesize\textsc{s}}} \\
\end{array}%
\hskip-3pt \right)\hskip-2pt,
\end{equation}
while the case of $\gamma=0$ can be obtained from (\ref{gen6}) by
changing the sign of $\mu_{\mbox{\footnotesize\textsc{a}}}$.

A small complex phase $\varepsilon\to 0$ of
$\gamma=\pi-\varepsilon$ produces the perturbation to
(\ref{gen6}), so that to the linear order in $\varepsilon$ it is
equal to
\begin{equation}\label{gen7}
    \mathcal{V}=
    \hskip-1pt
    \left(%
    \hskip-2pt
\begin{array}{ccc}
  0 & \mathrm{i}\mu_{\mbox{\footnotesize\textsc{a}}}\varepsilon & \hskip2pt0 \\[1mm]
  -\mathrm{i}\mu_{\mbox{\footnotesize\textsc{a}}}\varepsilon & 0 & \hskip2pt0 \\[1mm]
  0 & 0 & \hskip2pt0 \\
\end{array}%
\right)\hskip-2pt.
\end{equation}

Matrix (\ref{gen6}) can cause the hierarchy in both the masses and
mixings of fermion generations. Indeed, its eigenvalues are given
by
\begin{equation}\label{gen8}
    \begin{array}{rl}
      m_{1,2} = & \displaystyle
      \frac{1}{2}\bigg(\widetilde\mu_{\mbox{\footnotesize\textsc{a}}}+m_{\mbox{\footnotesize\textsc{s}}} \pm
      \sqrt{(\widetilde\mu_{\mbox{\footnotesize\textsc{a}}}+m_{\mbox{\footnotesize\textsc{s}}})^2+
      8\mu_{\mbox{\footnotesize\textsc{b}}}^2}\,\bigg),\\[3mm]
      m_3= & m_{\mbox{\footnotesize\textsc{x}}}+
      \mu_{\mbox{\footnotesize\textsc{a}}},
    \end{array}
\end{equation}
where $\widetilde\mu_{\mbox{\footnotesize\textsc{a}}}=
m_{\mbox{\footnotesize\textsc{x}}}-\mu_{\mbox{\footnotesize\textsc{a}}}$.
Setting
\begin{equation}\label{gen9}
    \mu_{\mbox{\footnotesize\textsc{b}}}^2\ll (\widetilde\mu_{\mbox{\footnotesize\textsc{a}}}
    +m_{\mbox{\footnotesize\textsc{s}}})^2
\end{equation}
at $\widetilde\mu_{\mbox{\footnotesize\textsc{a}}}>0$ and
$m_{\mbox{\footnotesize\textsc{s}}}>0$, we obtain
\begin{equation}\label{gen10}
    \begin{array}{rl}
      m_2\approx & \widetilde\mu_{\mbox{\footnotesize\textsc{a}}}+
      m_{\mbox{\footnotesize\textsc{s}}},
      \\[2mm]
      m_1\approx &\displaystyle
      -\frac{2\mu_{\mbox{\footnotesize\textsc{b}}}^2}
      {\widetilde\mu_{\mbox{\footnotesize\textsc{a}}}+m_{\mbox{\footnotesize\textsc{s}}}}.
    \end{array}
\end{equation}
Therefore, conditions
$m_{\mbox{\footnotesize\textsc{x}}}+\mu_{\mbox{\footnotesize\textsc{a}}}\gg
\widetilde\mu_{\mbox{\footnotesize\textsc{a}}}\gg
m_{\mbox{\footnotesize\textsc{s}}}$ lead to the hierarchy of
fermion masses
\begin{equation}\label{gen11}
    m_3\gg m_2\gg |m_1|.
\end{equation}
Two lighter generations are formed by superposition of two initial
states defining matrix (\ref{gen6}). The superposition is simply
the rotation with angle
$\theta_{\mbox{\footnotesize\textsc{c}}}^\prime$
\begin{equation}\label{gen12}
    \tan
    2\widetilde\theta_{\mbox{\footnotesize\textsc{c}}}=\frac{2\sqrt{2}
    \mu_{\mbox{\footnotesize\textsc{b}}}}{\widetilde
    \mu_{\mbox{\footnotesize\textsc{a}}}+m_{\mbox{\footnotesize\textsc{s}}}},
\end{equation}
which is the analogue of Cabibbo angle, since the electroweak
partners of fermion fields could have got the mass matrix with the
same texture, that leads to the similar mixing of initial states,
so that initially diagonal electorweak charge currents acquire the
mixing of two lighter generations with the angle given by the
difference of $\widetilde\theta_{\mbox{\footnotesize\textsc{c}}}$
parameters for two kinds of fields.

Thus, the realistic scenario suggests the texture with
\begin{equation}\label{gen13}
\begin{array}{rcl}
    m_{\mbox{\footnotesize\textsc{x}}}\sim\mu_{\mbox{\footnotesize\textsc{a}}}\sim
    \mu_{\mbox{\footnotesize\textsc{x}}}&\gg&
    |m_{\mbox{\footnotesize\textsc{x}}}-\mu_{\mbox{\footnotesize\textsc{a}}}|\gg
    m_{\mbox{\footnotesize\textsc{s}}},\\[1mm]
    |m_{\mbox{\footnotesize\textsc{x}}}-\mu_{\mbox{\footnotesize\textsc{a}}}|&\gg&
    \mu_{\mbox{\footnotesize\textsc{b}}},\\[2mm]
    \mu_{\mbox{\footnotesize\textsc{b}}}^2&\gg&
    |m_{\mbox{\footnotesize\textsc{s}}}-\mu_{\mbox{\footnotesize\textsc{a}}}|\,
    m_{\mbox{\footnotesize\textsc{s}}}.
\end{array}
\end{equation}
Such the hierarchy can be natural, since parameters
$m_{\mbox{\footnotesize\textsc{x}}}$ and
$\mu_{\mbox{\footnotesize\textsc{a}}}$ are determined by the
vacuum state with SUSY broken down at the scale
$\mu_{\mbox{\footnotesize\textsc{x}}}$, while
$m_{\mbox{\footnotesize\textsc{s}}}$ could be equal to zero in the
case of exact SUSY. Then, the only condition required is a small
mixing between the ordinary fermionic fields in the sectors with
broken and exact SUSY, that could serve as the definition of
\textit{ordinary} matter fields\footnote{We do not consider the
neutrino masses and mixing, here, since this problem requires a
more fine treatment because of an extremely low value of neutrino
mass scale caused by physics beyond the Standard Model.} in
contrast to the \textit{hidden} sector, wherein one could expect
$\mu_{\mbox{\footnotesize\textsc{b}}}\sim\mu_{\mbox{\footnotesize\textsc{x}}}$,
which breaks the hierarchy down, and it leads to hidden fermions
with masses of the order of
$\mu_{\mbox{\footnotesize\textsc{x}}}$.

At the limit of $\mu_{\mbox{\footnotesize\textsc{b}}}\to 0$, one
can expect the observation of almost massless generation of
ordinary fermions with respect to the scale of SUSY breaking, of
course. This fact is in agreement with the experimental data.

The correction to the symmetric case produces a small mixing of
heavy generation of ordinary matter with two lighter generations
as well as the violation of $\mathbb{CP}$-invariance due to
(\ref{gen7}). This correction can be treated perturbatively, if
appropriate matrix elements of $\mathcal{V}$ is much less than the
splitting between the levels, i.e. at $|\mathcal{V}|\ll\Delta E$,
that simply gives $\epsilon\ll 1$.

Similar features of generation structure can be exported to the
sector of fermion superpartners, i.e. scalar fields of sfermions.
Then, in the action quadratic versus the sfermions, the mass
matrix is composed by both squares of initial masses positioned at
the diagonal and non-diagonal mixing parameters. Since the
eigenvalues could turn out to be negative, the negative sign would
indicate the generation of sfermion condensates. Anyway, the
observational situation suggests that sfermions have no light
states analogous to the almost massless generation. Therefore,
sfermions should imitate the texture of hidden sector.

The question about gauge vector fields is more specific, since
such the fields acquire masses due to the higgs effect. We can
suppose that the observed mediators of gauge interactions couple
to the lightest generation of higgs field, while two more heavy
hidden generations of higgs scalars as well as, probably, gauge
fields can be discovered at the energy scale of SUSY breaking
down.

\section{Discussion and Conclusion\label{S-VI}}

In the present paper we have offered the ansatz for the low energy
modification of bare quadratic potential for the phenomenological
real scalar field, so that the correction parameterizes the energy
density given by zero-point modes of quantum fields due to the
supergravity relation between the superpotential and energy
density linear in the Newton constant $G$. So, the low energy
superpotential generates three local minima: one minimum
corresponds to superysmmetric flat vacuum-state, while two other
minima give the SUSY breaking down. To the same order in $G$, the
bare superpotential induces the bare quartic term of potential, so
that the actual potential taking into account all of bare and low
energy terms has the characteristic form with the barrier
separating the flat vacuum from two AdS-states. These initial
vacuum-states are not stationary, since the bubbles of AdS-vacua
in the flat vacuum generate the fluctuations inducing the mixing.
Putting the domain wall thickness equal to the inverse bare mass
by the order of magnitude and the low energy quartic coupling
equal to the bare value, we have determined the bare mass scale
$\widetilde m \sim 10^{12}$ GeV and quartic coupling $\lambda\sim
10^{-14}$.

Having took into account the mixing described phenomenologically, we
have found the stationary vacua represented by superpositions of
initial flat and AdS-states. So, we have got two AdS-states with
vacuum energy scale about the scale of SUSY breaking down, while the
single dS-state acquires the cosmological constant consistent with
the experimental data. Then, statically the offered potential
naturally gives the vacuum with the desired small value of
cosmological constant determined via the seesaw mechanism of mixing
in terms of tuned potential barrier.

Further, we have shown, that the same scalar field can serve as
the inflaton. The observations prefer for the quadratic term
dominance in the region of Planckian fields. This constraint has
been analyzed by means of quasi-attractor approach. Then, the
inhomogeneity of matter in the Universe is in agreement with the
quantum fluctuations of field during the inflation. The inflaton
decays into massless quanta in vicinity of potential barrier, the
height of which determines the temperature of preheating. This
decay can take place on the background of flat vacuum, only. Then,
the domain wall fluctuations transform the flat vacuum into the
stationary dS-state, while the stationary AdS-states remain beyond
the play.

The vacuum structure causes three fermion generations. We have
analyzed the textures of mass matrices. So, the hierarchies for
masses and mixing of the ordinary matter in the observable sector
can be consistently constructed, or otherwise heavy states in the
hidden sector and in the sector of superpartners for the ordinary
fields can be introduced.

Thus, the offered model of potential is suitable for the
description of three benchmarks: the naturally small cosmological
constant, inflation with further preheating stage driving to the
dS-state, and three fermion generations with appropriate
hierarchies and mixing.

In this scheme all of SUSY breaking effects as well as consequent
low energy condensates and phase transitions are accumulated in
the energy density of initial AdS-vacua, while the flat vacuum is
preserved from its influence due to the exact SUSY. The mixing of
flat and AdS-states is described phenomenologically in terms of
real scalar field. We have argued for the dynamical evolution to
the dS-vacuum. So, we have presented the natural mechanism of
driving to the observed cosmological constant, that is alternative
to the renormmalization group arguments developed in
\cite{ShapiroSola,ShapiroSola-2,Guberina,Shapiro:2004,Bilic,Sola,rev-CC}.

\section*{Acknowledgement}
This work was partially supported by grants of Russian Foundations for Basic
Research 09-01-12123 and 10-02-00061, Special Federal Program ``Scientific and
academics personnel'' grant for the Scientific and Educational Center
2009-1.1-125-055-008, ant the work of T.S.A. was supported by the
Russian President grant MK-406.2010.2.
%

\end{document}